\documentclass[final,5p,times,twocolumn]{elsarticle} 
\usepackage{lineno,hyperref}

\journal{Elsevier}

\usepackage{gensymb}
\usepackage{graphicx}
\usepackage{subcaption}
\usepackage{amsmath}
\usepackage[capitalise]{cleveref}
\usepackage{upgreek}

\usepackage{pdfpages}
\usepackage{setspace}
\usepackage{comment}

\usepackage{epstopdf}
\AppendGraphicsExtensions{.tif}

\begin{document}

\begin{frontmatter}

\title{Depth-dependent charge collection profile of pad diodes}

\author[1]{M. Hajheidari \corref{cor1} }

\author[1]{E. Garutti}
\author[1]{J. Schwandt}
\author[2]{A. Ebrahimi}

\address[1]{Institute for Experimental Physics, University of Hamburg, Luruper Chaussee 149, 22761 Hamburg, Germany}

\address[2]{Paul Scherrer Institut, Villigen, Switzerland}

\cortext[cor1]{Corresponding author: mohammadtaghi.hajheidari@desy.de}

\begin{abstract}
The collected charge of two pad diodes is measured along the diode width using a $5.2 \, \text{GeV}$ electron beam at the DESY II beam test facility. The electron beam enters parallel to the readout electrode plane and perpendicular to the edge of the diode. The position of the electron beam is reconstructed by three planes of an EUDET-type telescope. An in-situ procedure is developed to align the diode surface parallel to the electron beam. The result of these measurements is the charge collection efficiency profile as a function of depth for each diode. 
For a non-irradiated diode, the charge profile is uniform as a function of the beam position for bias voltages above the full-depletion, as expected. For the irradiated diode, the charge profile is non-uniform and changes as a function of bias voltage.

\end{abstract}


\end{frontmatter}


\section{Introduction}

\par Transient Current Technique (TCT) is a well-known method to study irradiated silicon sensors. In one flavour of this technique so-called Top-TCT, the Device Under Test (DUT) is illuminated by $\alpha$-particles \cite{kramberger2002effective,
kramberger2002determination,brodbeck2000new} or red light and infra-red (IR) laser \cite{kraner1993use,beattie1999carrier,
fink2006characterization}, and the induced current transient is recorded. One can determine the collected charge by integrating this transient. The result of this measurement is a quantity called "Charge Collection Efficiency" or CCE defined as the ratio between the collected charge by the electrodes and the produced charge inside the sensor.  

\par Due to the short absorption depth of $\alpha$-particles and red light in silicon, all electron-hole pairs are generated close to the surface of the diode. Therefore, only one of the charge carrier types (electrons or holes) is drifting through the bulk region while the other is collected immediately at the implant. The induced signal in the electrodes of the diode is dominated by the charge carrier type which is drifting through the entire bulk and the other type has a minor contribution to the signal. As a result, one can study the trapping times of electrons and holes separately using this technique for irradiated sensors. Using IR light, one can obtain the overall CCE of irradiated samples with contribution from both electrons and holes.

\par Top-TCT with alpha particle and red -light laser is limited for studying highly irradiated pad diodes. The estimated CCE values from these measurements are very sensitive to possible inactive layers in the sensor implants. There is a field-free region in the diode implant where the generated electron-hole pairs can be trapped before they diffuse to the active region of the diode (the region with a non-zero electric field) \cite{scharf2015measurement, scharf2018radiation}. Simulation of a non-irradiated diode shows that this region remains field-free independent of the applied voltage. As a result, the measured CCE values with the Top-TCT method using alpha-particles or red light underestimate the actual CCE in the active region of irradiated diodes and are affected by the change of the thickness of the non-depleted layer, depending on the irradiation fluence. 

\par In another flavour of this technique called "edge-TCT" or E-TCT, the sensor is illuminated by focussed IR laser light from the edge \cite{Kramberger:2010tem}. By scanning the light along the edge, one can obtain information on the charge collection as a function of depth. This method has been used to study strip sensors to extract their charge and electric field profiles \cite{mandic2015edge, mandic2013tct, klanner2020determination}. An issue with this technique is that the waist radius of the light spot changes as it travels through the sensor and the beam becomes defocused. As a result, it can only be used for segmented sensors (strips and pixels) where the charge produced below a narrow strip is collected and read out. This technique cannot be used for measuring pad diodes where there is only one readout electrode. Another issue is that it is not straightforward to normalize E-TCT measured charges to an absolute value, as the exact number of electron-hole pairs generated by light in the sensor is not known. In addition, it has been shown that the absorption length of infrared light changes after irradiation \cite{scharf2020influence}. Therefore, a direct comparison between irradiated and non-irradiated detectors is not straightforward.

\par This work uses an electron beam ($5.2 \, \text{GeV}$) for edge-on measurements on pad diodes. Charge collection profiles as a function of depth have been measured for pixel detector at grazing angles \cite{chiochia2005simulation}. For pad diodes, the edge-on method was introduced in \cite{gorivsek2016edge}. In that work, pions with an energy of $120 \, \text{GeV}$ were used to study $300 \,\upmu \text{m}$ thick pad diodes. However, no results about the charge collection of the studied diode were shown in that paper. Moreover the required alignment accuracy was not discussed and no alignment procedure was presented. It should be noted that the alignment has to be performed on-line so that the data are taken with the sensor aligned to the beam.  

\par The main aim of this study is to obtain the position-dependent charge collection perpendicular to the electrode plane. The Charge Collection Distance (CCD), or the mean distance in which charge carriers drift inside the sensor before trapping, can be estimated using the results of these measurements. The advantages of this technique compared to the Edge-TCT with IR light are: 
\begin{itemize}
\item Radius and direction of the beam do not change as it travels through the sensor. The estimated deflection angle of an electron beam with an energy of $5.2 \, \text{GeV}$ after travelling in $5.0 \, \text{mm}$ of silicon is $0.53 \, \text{mrad}$ \cite{lynch1991approximations}. This number is well within the uncertainty of the measurements.  Therefore, the method can be used to study pad diodes. 

\item An absolute normalization is possible as the energy loss of the beam in the sensor, i.e.\ average $dE/dx$, is well known. 

\item The energy lost of the electron beam by charge deposition does not change after irradiation. Therefore, one can directly compare the results of the non-irradiated and irradiated diodes together. 
\end{itemize}

The disadvantages of this technique are:
\begin{itemize}
\item Due to the large capacitance of pad diodes, rise times of induced signals on the electrodes are in a order of a few nanoseconds. As a result, it is not straightforward to extract information about charge carrier velocity and electric field from the transients. 

\item The alignment procedure proposed in this paper is limited for highly irradiated sensors where the charge collection profile is not uniform as a function of depth. 
\end{itemize}
The paper is structured in the following way. After introducing the experimental setup, an in-situ alignment is proposed to find the minimal angle between the electron beam and the diode surface. In the result section, the  collected charge profile for the no-irradiated is shown for different bias voltages and beam energies. In addition, the collected charge profile of an irradiated diode is shown as a function bias voltage.

\section{Experiment Setup}
\label{Experiment}

\par
\begin{figure}[ht]
  \centering 
     \includegraphics[width=1.0\linewidth]{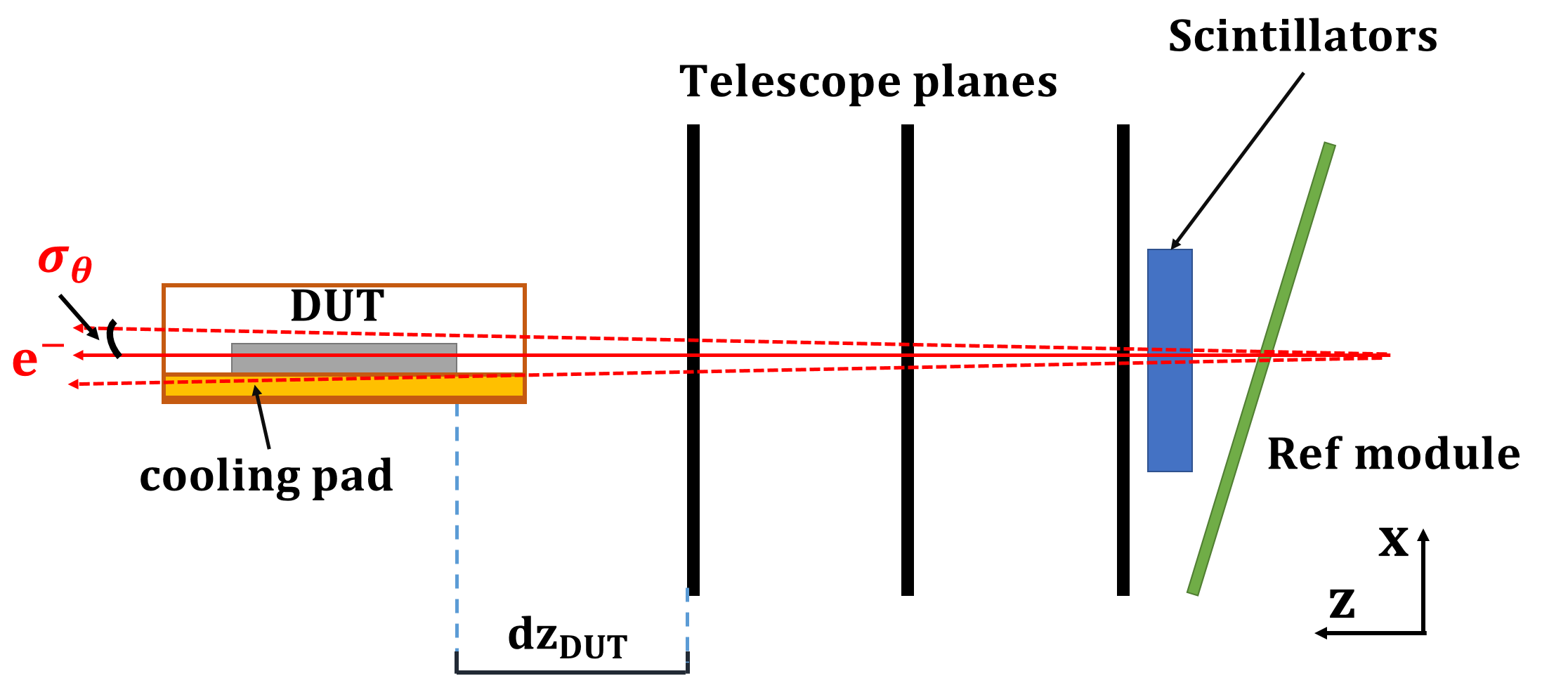}
    \caption{Schematic view of the beam test setup used for edge-on measurements. In this sketch, DUT (the Device Under the Test) is the pad diode, Ref module is the timing reference module, Scintillatiors provide the readout trigger, and Telescope planes reconstruct the tracks. The divergence of the electron beam is shown with two dash lines ($\sigma_{\theta}$). The distance between the third plane of the telescope and the DUT is shown with $dz_{dut}$.} 
     \label{setup}
 
\end{figure}

 The measurements are performed in at Deutsches Elektronen-Synchrotron (DESY) test beam facility. In this facility, electron/positron beams with energy in a range of $1-6 \, \text{GeV}$ are available \cite{diener2019desy}. For this work, the electron beam with an energy of $5.2 \, \text{GeV}$ is chosen. A schematic view of the setup used for these measurements is shown in \cref{setup}. 

\par The tracks are reconstructed using three planes of an EUDET-type telescope. Each plane of the telescope consists of a MIMOSA 26 sensor \cite{jansen2016performance}. This is a monolithic pixel sensor produced in $350 \, \text{nm}$ CMOS technology. The pixel size is $18.4 \times 18.4 \, \upmu  \text{m}^2$ and the thickness of the sensor is $54.5 \pm 3.6 \, \upmu \text{m}$. Each sensor contains 1152 columns and 576 rows which cover an area of $21.2 \times 10.6 \, \text{mm}^2$. The readout of the sensors is binary. The estimated intrinsic hit resolution of a single plane is $3.24 \pm 0.9 \, \upmu \text{m}$ \cite{hu2010first}. The threshold of the telescope sensors is set to $4 \, \text{mV}$ corresponding to $\sim 70 \, \text{e}^{-}$.

\par The integration time of the MIMOSA 26 planes is relatively long ($115 \, \upmu\text{s}$). Therefore, a time reference is needed to select a subset of tracks that are contained in one readout cycle. The time reference is provided by a CMS phase 1 pixel module. This module contains a silicon sensor with a pixel size of $100 \times 150 \, \upmu\text{m}^2$ and a PSI46dig readout chip. The internal clock of the readout chip is $40 \, \text{MHz}$ (correspond to $25 \, \text{ns}$ readout cycle). 

\par The trigger signal for readout of the telescope, timing reference module, and DUT is provided by two orthogonally mounted scintillators coupled to Photo Multiplier Tubes (PMTs). The scintillators are placed in front of the telescope planes. The coincidence signal between them provides a triggering area of $8 \times 3 \, \text{mm}^2$. The output of the PMTs is fed to a device called Trigger Logic Unit (TLU) \cite{cussans2009description}. The TLU provides the trigger signal through a logic AND between signals of two scintillators.

\par
\begin{figure}[ht]
  \centering 
  
    \includegraphics[width=1\linewidth]{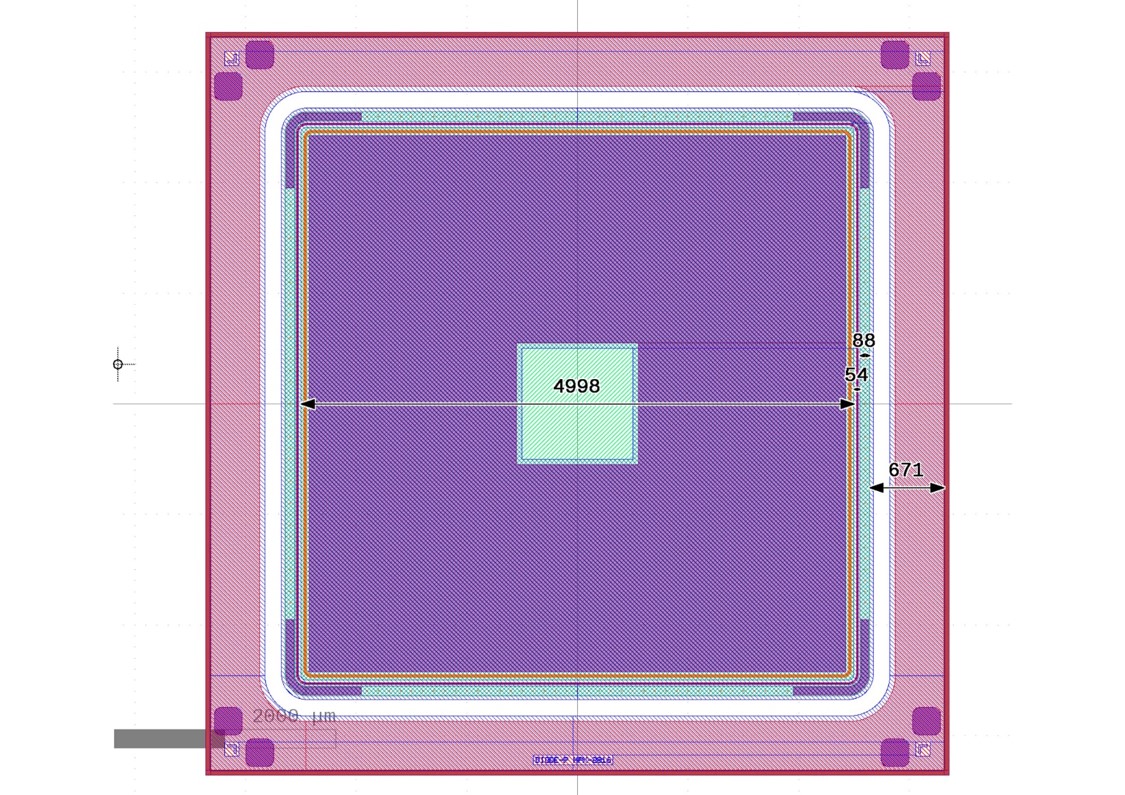}
    \caption{Top view of the diode (dimensions in $\upmu \text{m}$). }
      \label{diode_top}

  \end{figure}

 The diodes in this work are $p$-type ($n^+pp^+$configuration) with an active thickness of 150 $\: \upmu \text{m}$. The active area of diodes is $\approx 25 \, \text{mm}^2$. \cref {diode_top} show the top view of the studied diodes. One diode is irradiated with $23 \, \text{MeV}$ protons to a $1 \, \text{MeV}$ neutron equivalent fluence $\Phi_{eq}$ of  $2\times 10^{15} \, \text{cm}^{-2}$. For the calculation of $\Phi_{eq}$, a hardness factor $\kappa = 2.2$ is used \cite{allport2019experimental}. From the capacitance-voltage measurement, the depletion voltage of the non-irradiated diode is determined to be around $75 \, \text{V}$ and the doping density of the bulk region $4.5 \times 10^{12} \, \text{cm}^{-3}$ . The guard ring of the diodes is floating during the measurement.

\par A Rohde \& Schwarz oscilloscope with an analog bandwidth of $4 \, \text{GHz}$ and a sampling rate of $20 \, \text{GS/s}$ records the transients. A Femto HSA-X-40  amplifier with a bandwidth of $2.0 \, \text{GHz}$ and a nominal gain of $100$ amplifies the transient \cite{Femto}. The trigger is provided externally through the TLU. The external triggering makes it possible to record all transients without zero suppression. 

\par To calculate the collected charge, transients are integrated in a time window (gate). The charge is calculated as: 

\begin{equation}
{Q}=\int_{{t}_0}^{{t}_1} \frac{{U(t)}}{G \cdot {R}_{{L}}} dt
\label{E1}
\end{equation}
$U(t)$ is the transient after the baseline correction and $R_L$ is the input impedance of the oscilloscope ($50 \, \ohm $) and $G$ is the nominal gain of the amplifier (100). For this study, a gate width of $30 \, \text{ns}$ is chosen.

\par To measure the transients, the diode is mounted on a PCB which provides the electrical connectors. The parallel beam can produce showers inside the PCB. These showers can produce particles which deposit energy in the DUT. To cancel the effects of these showers on the response of the diode, two spacers are inserted between the diode and the PCB. These spacer are metal bars with a width of $w = 1 \, \text{mm}$ and a thickness of $d= 0.5 \, \text{mm}$.

\section{Online alignment}
\label{alignment}
\par The method proposed in this paper only works if tracks are parallel to the diode surface. To ensure this condition, an "online alignment" procedure is used before data taking. It is important to point out that this procedure has to be done online and before one can start taking data and cannot be corrected offline.  

\par In this procedure, the mean collected charge is measured as a function of angle with fine steps ($0.1^ \circ$ or $ \approx 1.7 \, \text{mrad}$). The minimum step size of the rotation stage is $0.01^\circ$. The minimum or "zero" angle is defined as the angle where the measured charge is maximum. If tracks and diode surface are not parallel, a part of the tracks leaves the diode through the top or bottom plane and therefore deposit less charge than tracks that cross the whole diode ($5 \, \text{mm}$). As a result, the average charge is maximal when the angle between the tracks and the diode is minimal. The important assumption in this procedure is that the collected charge is uniform as a function of depth. 
\par The uncertainty of the position reconstruction due to the angular spread of the beam ($\sigma_{\theta}$) is given by $d \cdot \sigma_{\theta}$, where $d$ is the length of the diode which is $5 \, \text{mm}$ for this study. This quantity should be kept smaller than the resolution $\sigma_{x,beam}$. The uncertainty due to the step size of the angle scan ($\Delta \theta$), which is given by $d \cdot \Delta \theta$, should be smaller than the resolution $\sigma_{x,beam}$ as well.
\par In \cite{jansen2016performance}, the telescope resolution was calculated as a function of the DUT-to-telescope distance (i.e.\ $dz_{DUT}$ in \cref{setup}). By assuming a resolution around 10 $\upmu \text{m}$, the maximum value for $\sigma_{\theta}$ and $\Delta \theta$ is estimated to be $2.0 \, \text{mrad}$. In the offline analysis, a cut is applied on the  slope of  reconstructed track to select the ones with a slope less than $\pm 1.0 \, \text{mrad}$. The step size of the angle scan is chosen $0.10^\circ$ or $1.7 \, \text{mrad}$.
\par
\begin{figure}[ht]
  \centering 
  \begin{subfigure}[b]{0.8\linewidth}
 
    \includegraphics[width=1\linewidth]{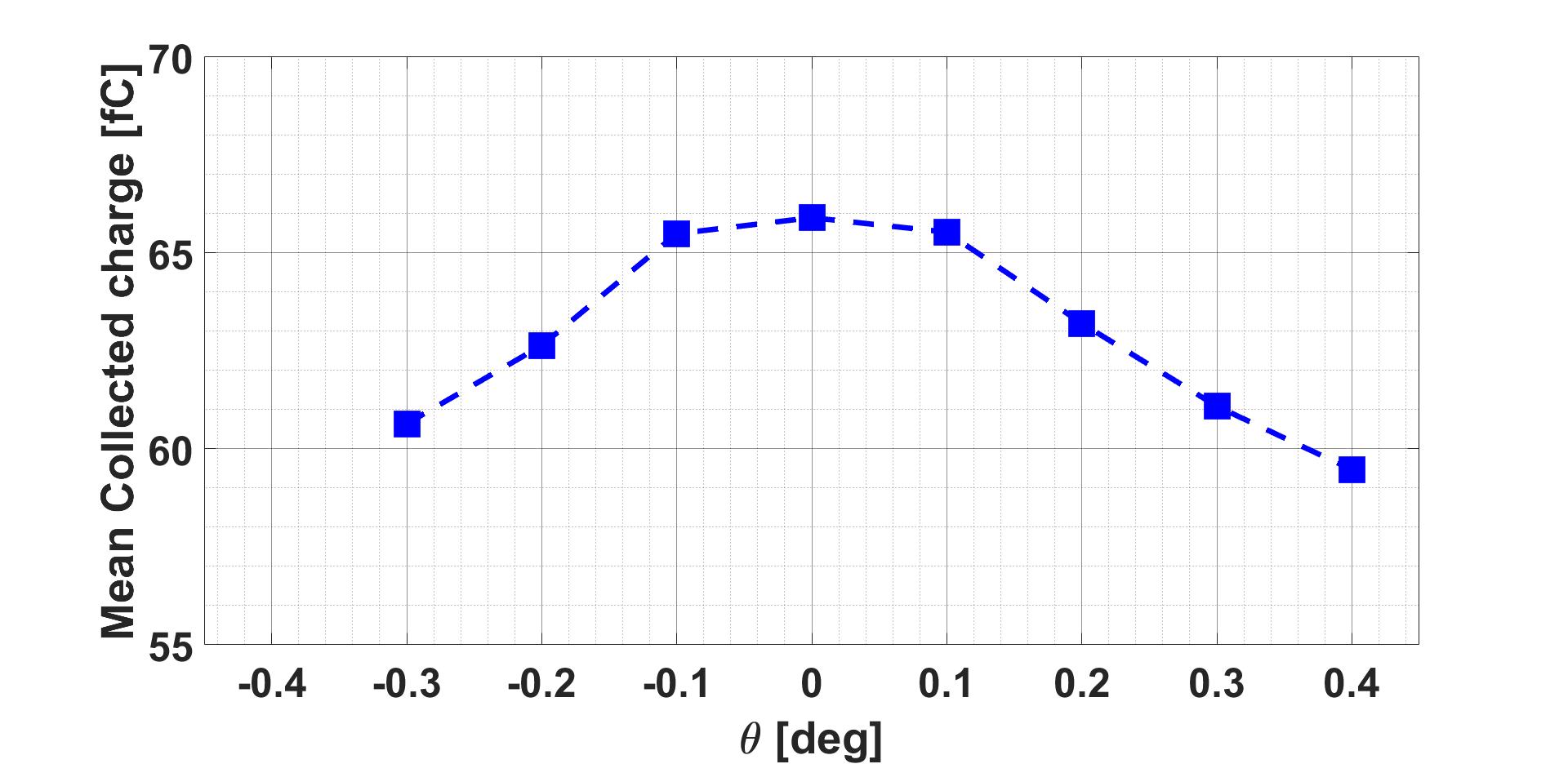}
    \caption{Mean collected charge as a function of the angle ($\theta$) for the non-irradiated diode. The plot is obtained for transients with amplitude higher than $20 \, \text{mV}$.}
     \label{charge_mean}
  \end{subfigure}
  \begin{subfigure}[b]{0.8\linewidth}
 
    \includegraphics[width=1\linewidth]{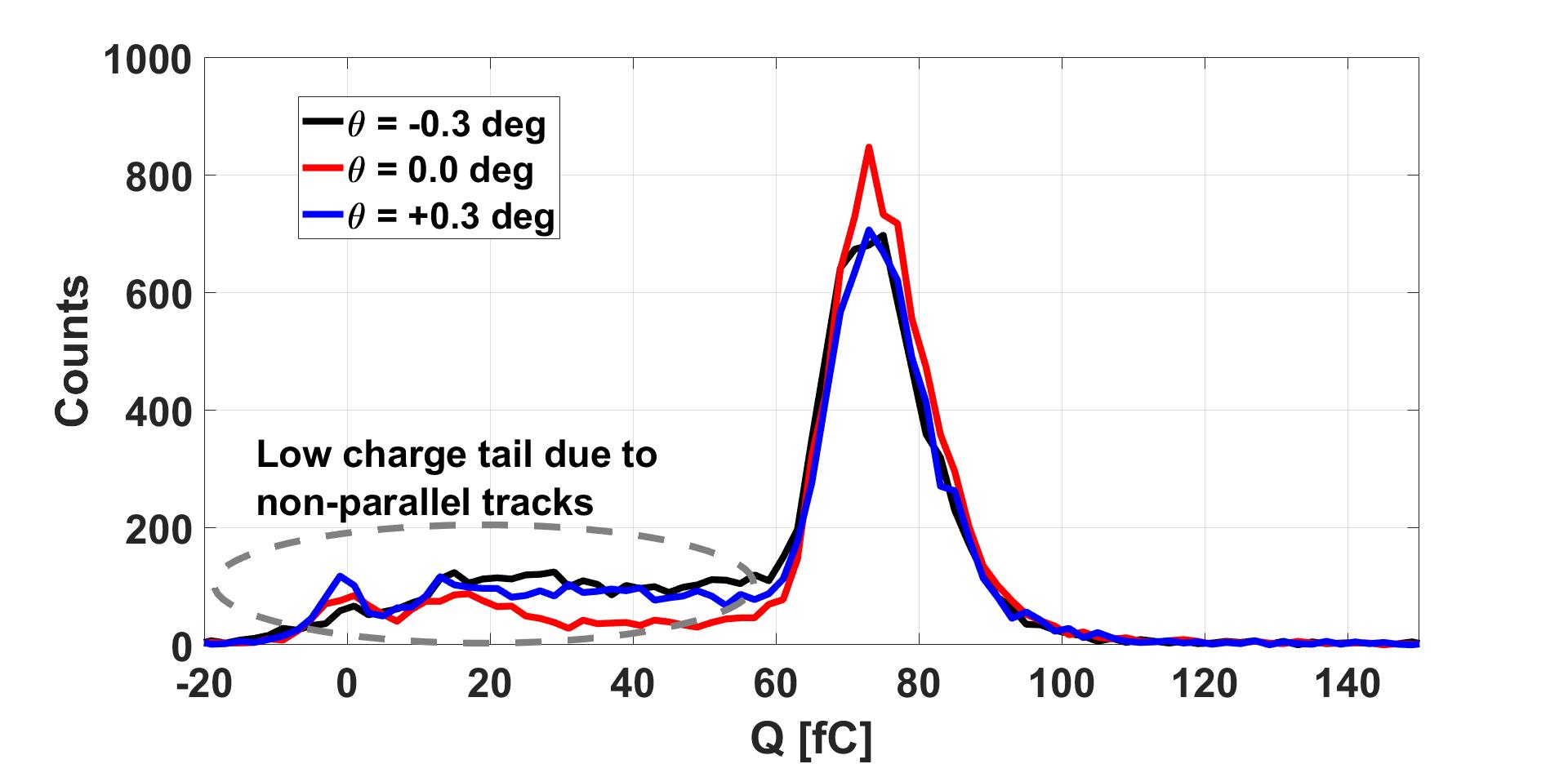}
    \caption{Collected charge distribution at three angles for non-irradiated diode.}
    \label{charge_dist}
  \end{subfigure}
  \caption{Results of the online alignment procedure needed to adjust the angle of the diode w.r.t. the beam before starting data taking. As a result of the scan, the angle for which the mean collected charge is maximum is defined to be $\theta = 0^\circ$.}
  \label{anglescan}
\end{figure}

 \cref{charge_mean} shows the average collected charge of the non-irradiated diode as a function of angle ($\theta$). The measurement is done at a bias voltage of $120 \, \text{V}$ and at room temperature. For each angle, 20,000 triggers are collected. To obtain this plot, an offline threshold of $20 \, \text{mV}$ is applied to the recorded transients. \cref{charge_dist} shows the charge distributions for three angles. One can notice, for non-zero angles, the charge distributions show a tail below the peak around $75 \, \text{fC}$ towards the lower charges. This tail corresponds to the tracks which partially crossed the diode and it is minimal when $\theta$ is minimal. 

\par The procedure exploits a simple geometrical relation between the track length inside the DUT and the angle of incidence. The advantage of using this procedure is that no track reconstruction was required. This procedure had to be repeated once the diode is changed or the cooling box is dismounted. 

\section{Results for the non-irradiated diode}
\label{Non-irrad}
 \par After finding the minimum angle, the final data is taken. The beam tracks are reconstructed and projected to the surface of the diode facing the beam. To gain maximum acceptance, the centering of the diode with respect to the trigger area is checked.

\par
\begin{figure}[h!]
  \centering
    \includegraphics[width=1\linewidth]{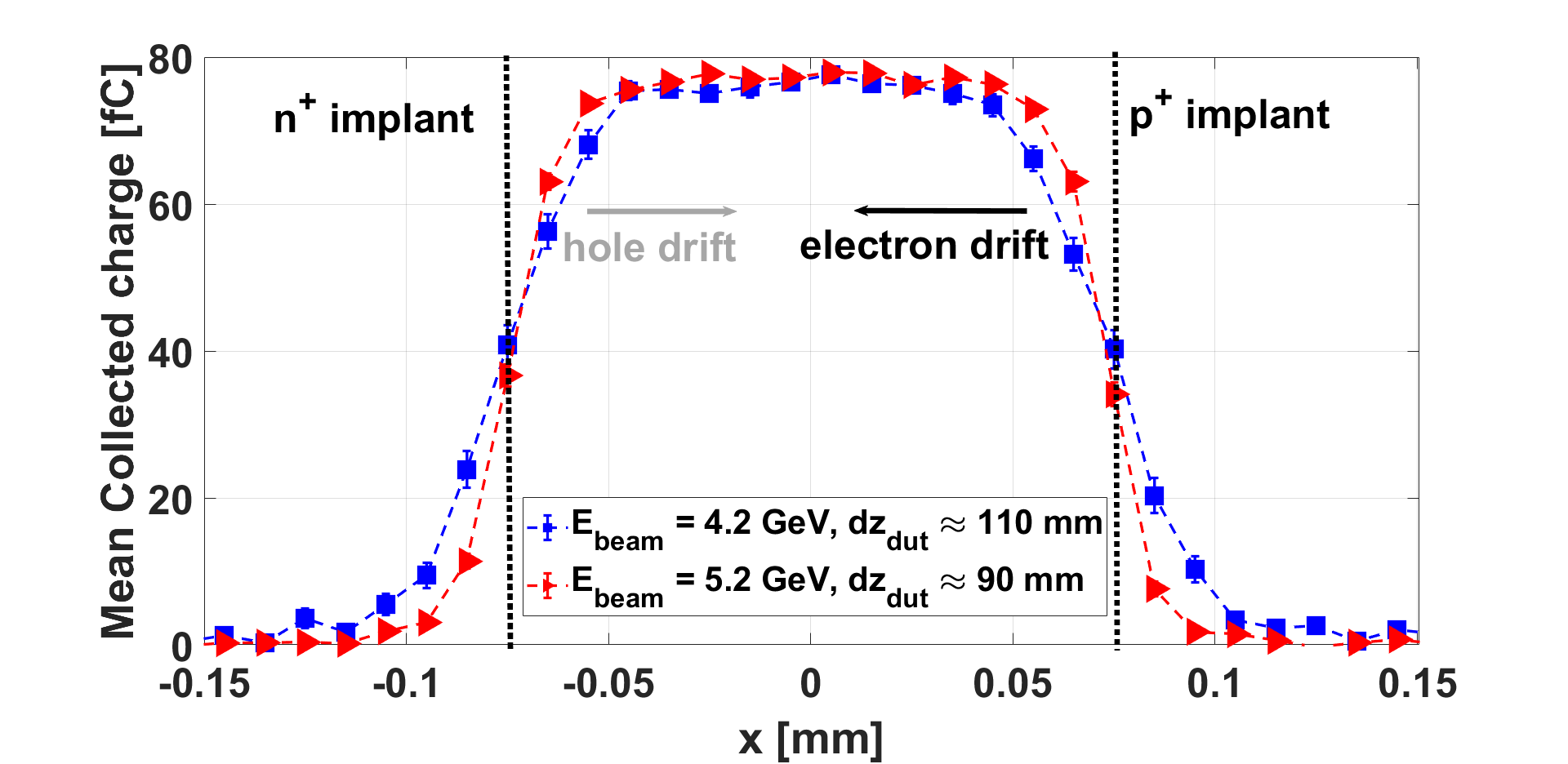}
    \caption{Charge profile of the non-irradiated diode at bias voltage of $120 \, \text{V}$.}
     \label{Nonirradvsx}
     \end{figure}

 \cref{Nonirradvsx} shows the charge profile as a function of depth for the non-irradiated diode at a bias voltage of $120 \, \text{V}$ and at room temperature. For comparison, the measurement is repeated at a lower beam energy of $4.2 \, \text{GeV}$ and different diode-to-telescope distance ($dz_{dut}$). One can see by increasing $E_{beam}$ and decreasing $dz_{dut}$, the profile at the edges become sharper and the spatial resolution of the measurement improves.

 \par
\begin{figure}[h!]
   \includegraphics[width=1\linewidth]{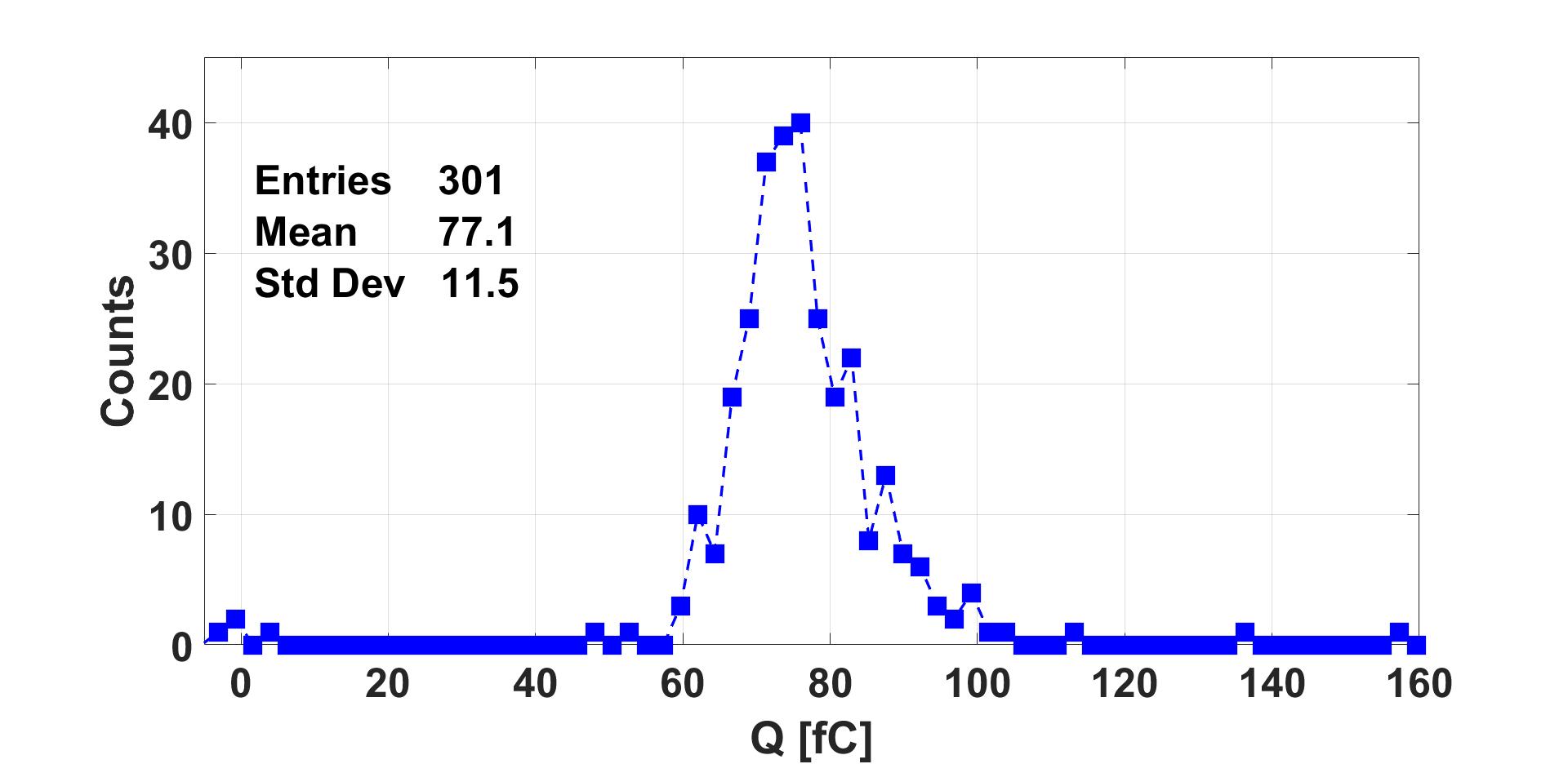}
    \caption{The charge distribution for a central bin with$-10 \, \upmu \text{m}<x<+10 \, \upmu \text{m}$.}
    \label{Nonirradtwobins}
    \end{figure}

 \cref{Nonirradtwobins} shows the charge distribution for bins with $ -10 \, \upmu \text{m}<x<+10 \, \upmu \text{m}$. Using a GEANT4 simulation, the mean deposition energy of $5.0 \, \text{GeV}$ electron beam in $5 \, \text{mm}$ of silicon is estimated to be $1.767 \, \text{MeV}$. One can convert the deposited energy to the deposited charge by taking into account the ionisation energy of silicon, $3.64 \, \text{eV}$ and the elementary charge:
 
\begin{align*} 
Q_{\text{deposit}}=E_{\text{deposit}} \cdot\frac{1.6 \times 10^{-19} \, \text{C} }{3.64 \, \text{eV} }=77.6 \text \, \text{fC}
\end{align*}

This value is in a good agreement with the measured absolute charge shown in \cref{Nonirradtwobins}. 

\par
\begin{figure}[h!]
  \centering 
  
    \includegraphics[width=1\linewidth]{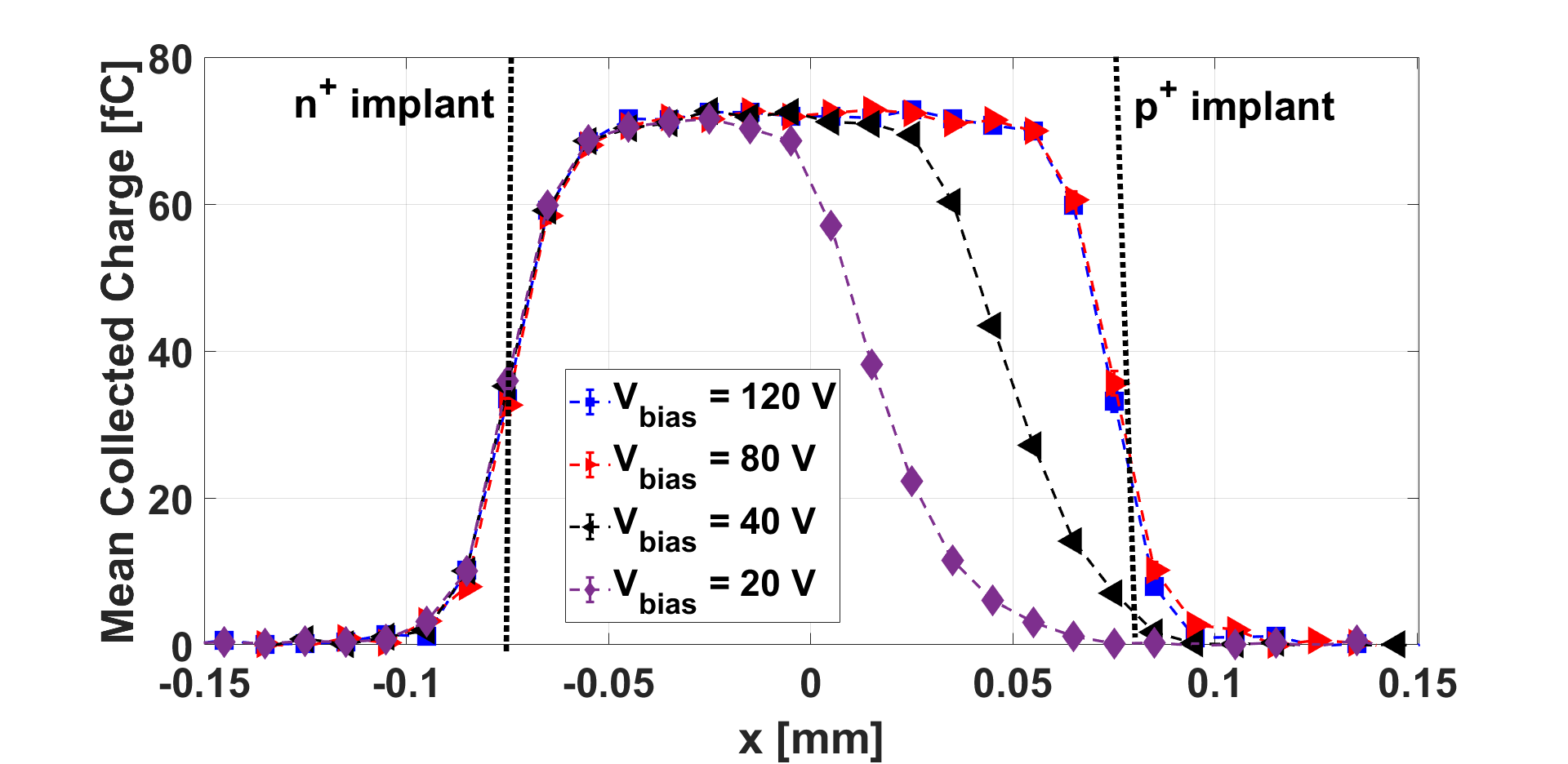}
    \caption{Charge collection profiles of non-irradiated diode at few bias voltages.}
     \label{Nonirradbiasscan}
  \end{figure}

 The measurements are repeated for bias voltages in the range of $20-120 \, \text{V}$. \cref{Nonirradbiasscan} shows the charge profiles obtained from these measurements. One can see that for bias voltage above $80 \,  \text{V}$, the profile remains unchanged. At bias voltage of $40 \, \text{V}$ and $20 \, \text{V}$, the charge collection is zero in the region near the $p^{+}$ implant. This can be explained by the fact that the electric field expands from the $n^{+}$ to the $p^{+}$ implant. At bias voltages below the full depletion voltage, $75 \, \text{V}$, the electric field is zero near the $p^{+}$ region and results in a region with zero charge collection. 

\par One can estimate the active thickness of the diode from the profiles shown in \cref{Nonirradbiasscan} by fitting the following function to the data:
\begin{equation}
F(x) = A\cdot \int_{x}^{x+dx}\left ( erf \Big[\frac{x'-{\mu}_1}{\sqrt{2}\sigma}\Big]-erf \Big[\frac{x'-{\mu}_2}{\sqrt{2}\sigma}\Big]\right)\,dx'
\end{equation}
The active thickness is estimated as $\mu_2-\mu_1$.

\par For bias voltages of $40 \, \text{V}$ and $20 \, \text{V}$, the active thickness is estimated to be $115 \, \upmu \text{m}$ and $83 \, \upmu \text{m}$, respectively.  The depletion depth of the diode at a bias voltage of $V_{bias}$ can be calculated using the following equation: 

\begin{equation}
{d} = \sqrt{{\frac{2 \epsilon_s \epsilon_0 V_{bias}}{e N_A}}}
\label{E2}
\end{equation}

where $e$ is the elementary charge, $\epsilon_0$ the vacuum permittivity ($8.85 \times 10^{-12} \, \text{F} \cdot \text{m}^{-1}$), $\epsilon_s$ the relative permittivity of silicon ($11.7$) and $N_{A}$ the doping concentration of bulk region ($4.5 \times 10^{12} \, \text{cm}^{-3}$). The calculated depletion depth from \cref{E2} at bias voltages of $40 \, \text{V}$ and $20 \, \text{V}$ is $107 \, \upmu \text{m}$ and $76 \, \upmu \text{m}$, respectively. Compared to the estimated active thickness from the measurements, the calculated depletion depths are slightly lower. This can be explained by taking into account the effect of the diffusion of charge carriers from non-depleted region to the active region. Due to this effect, the thickness of the region with non-zero charge collection is slightly larger then the depletion depth. 

\section{Results for irradiated diodes}
\label{Irr}
\par
\begin{figure}[h!]
  \centering
 
    \includegraphics[width=1\linewidth]{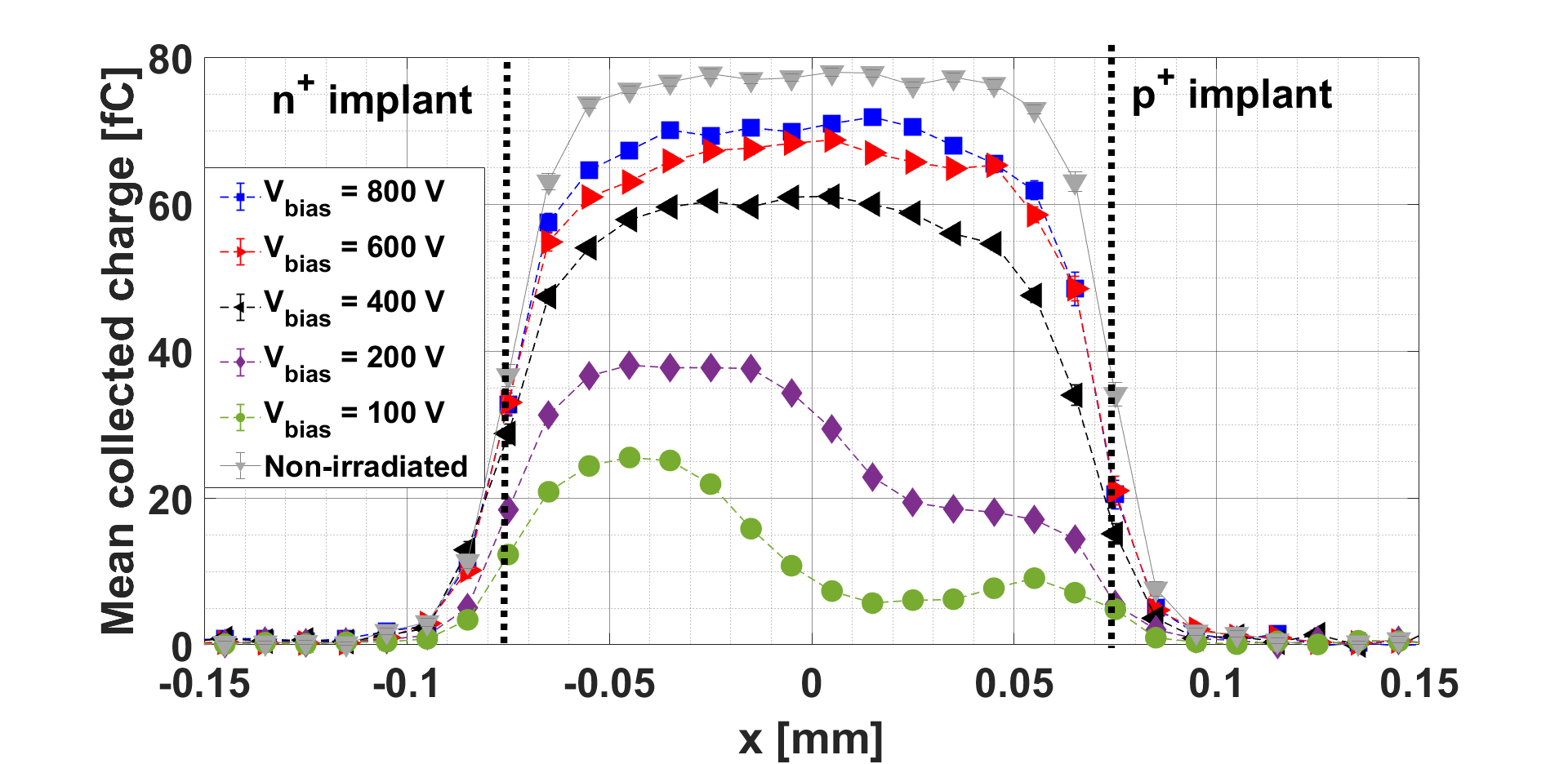}
    \caption{Charge collection profiles of non-irradaited diode at bias voltage of $120 \, \text{V}$ and irradiated diode ($\Phi_{eq} = 2 \times 10 ^{15} \, \text{cm}^{-2}$) at several bias voltages.}
     \label{irradvsx_2e15}
  
\end{figure}

 The measurements explained in \cref{Non-irrad,alignment} are repeated for an irradiated diode. In order to keep the leakage current of the diode low, it is mounted in a cold box. The box is cooled down using a circulation chiller (operating at $ - 20~^\circ \text{C}$)  and two Peltier elements. The exact temperature inside the box is not measured. \cref{irradvsx_2e15} show the charge profiles for the irradiated diode at $\Phi_{{eq}} = 2 \times 10 ^{15} \, \text{cm}^{-2}$ at different bias voltages, along with the charge profile of the non-irradiated diode at the bias voltage of $120 \, \text{V}$.

\par From these figures, one can see that the charge profile is not uniform as a function of depth, especially at low bias voltages. For both diodes, the collected charge near the $n^+$ implant, which is dominated by holes, is higher than at the $p^+$ implant. These results are in agreement with previously published results on velocity and electric field profiles of irradiated strip sensors \cite{kramberger2014modeling, klanner2020determination}. In these references, one can see high electric fields near $n^+p$ and $pp^+$ implants and a lower electric field in the center of the sensor. Moreover, one can see a more uniform electric field at high bias voltages which is in agreement with the measured collected charge profile presented in this paper. At the highest bias voltage, a reduced CCE is seen which is a consequence of trapping of charge carriers. 

\section{Conclusion}
\par In this paper, we demonstrate the feasibility of using an electron beam entering the side of a silicon pad diode to obtain the depth-dependence charge collection efficiency profile. We introduce an online method to find the minimum angle between the diode and electron beam.

\par It is shown that the charge collection profile of a non-irradiated pad diode is uniform as a function of depth for bias voltages above full depletion. The profile of irradiated diode, however, is not uniform, especially for lower bias voltages. For the highest applied bias voltage, $800 \, \text{V}$, the CCE is less than 1 for the investigated irradiated diode. 

\par Qualitatively, it is observed that the shape of measured charge collection profiles in this paper is in agreement with previously published measurements of the electric fields for irradiated strip sensors. 

\par It was shown that the method can be further optimized. The resolution can be improved by minimizing the distance between the telescope and the DUT and increasing the beam energy. The efficiency of data selection can be improved by optimizing the geometrical match between the DUT and the trigger scintillator.

\section{Acknowledgements}  

\par The authors acknowledge support by the BMBF, the German Federal Ministry of Education and Research. This work was funded by the Deutsche Forschungsgemeinschaft (DFG, German Research Foundation) under Germany’s Excellence Strategy -- EXC 2121 ``Quantum Universe'' -- 390833306.

\par The measurements leading to these results has been performed at the Test Beam Facility at DESY Hamburg (Germany), a member of the Helmholtz Association (HGF).

\bibliography{Ref}
\end{document}